\documentclass[aps,prb,twocolumn,groupedaddress]{revtex4} 
\usepackage{graphicx}% Include figure files  

\begin{document}  
\newcommand{\be}{\begin{equation}} 
\newcommand{\ee}{\end{equation}} 
\newcommand{\bea}{\begin{eqnarray}} 
\newcommand{\eea}{\end{eqnarray}} 
\newcommand{\nt}{\narrowtext} 
\newcommand{\wt}{\widetext}  

\title{On electron (anti)localization in graphene} 

\author{D. V. Khveshchenko} 

\affiliation{Department of Physics and Astronomy, University of North Carolina, Chapel Hill, NC 27599}  

\begin{abstract} 
We discuss localization properties of the Dirac-like 
electronic states in monolayers of graphite.  In the framework of 
a general disorder model, we identify the conditions under which 
such standard localization effects as, e.g., the logarithmic temperature-dependent 
conductivity correction appear to be strongly suppressed, as compared to 
the case of a two-dimensional electron gas with parabolic dispersion, in 
agreement with recent experimental observations.  
\end{abstract}  

\maketitle 

After several decades of predominantly application-driven studies, 
graphene has finally been recognized as a unique example of the 
system of two-dimensional fermions with a linear dispersion and 
pseudo-relativistic kinematical properties. The recent advances 
in microfabrication of graphitic samples that are only a few carbon layers thick 
\cite{geim,kim} have made it possible to test the early theoretical 
predictions of the anomalous properties of this system \cite{semenoff}. 

The most striking experimental observation up to date  was that of 
an anomalous quantization of the Hall conductivity \cite{geim,kim,gusynin} 
which is characteristic of the pseudo-relativistic nature of the quasiparticle 
excitations in this system. The other properties manifesting this peculiar 
single-particle kinematics have been revealed by magnetotransport measurements, 
including the ${\sqrt B}$ dependence of the energies of the (non-equidistant) 
Landau levels and the intrinsic $\pi$-shift of the  phase of the Shubnikov-de Haas oscillations
\cite{geim,kim,kopelevich}.   

In a (nearly) degenerate semimetal such as graphite, the Coulomb interactions 
are expected to play an important role, largely due to their poor screening \cite{guinea}. 
Besides, any interplay between the Coulomb interactions and disorder is likely to further  
modify the behavior of the idealized (clean and non-interacting) Dirac fermion system. 
Thus far, however, experiment has not yet provided a direct evidence of any 
interaction-induced phenomena, some of which were predicted in the recent years \cite{guinea,dvk}. 

Also, contrary to the situation in the conventional 2DEG, the data of 
Refs.\cite{geim,kim} do not seem to manifest any pronounced weak-localization effects, either. 
Motivated by these observations, in the present work we set out to study electron 
localization in graphene. Although this topic has already attracted some recent attention 
\cite{ando}, we shall demonstrate that the analysis of Ref.\cite{ando} is neither 
complete, nor (curiously enough, considering that the assertions made in 
Ref.\cite{ando} are mostly correct) do 
the arguments presented in that work appear to be technically sound.    

The electronic band structure of graphene is characterized by the presence of a 
pair of inequivalent nodal points at the wave vectors ${\vec K}_{1,2}={2\pi/9a}
(\pm{\sqrt 3},3)$ where $a$ is the lattice spacing. At these two and the four other points in the Brillouin zone
obtainable from ${\vec K}_{1,2}$ with a shift by one of the reciprocal lattice 
vectors ${\vec Q}_{1,2}={2\pi/3a}({\sqrt 3},\pm 1)$, the valence and conduction 
bands touch upon each other as a pair of opposing cones with the opening angle given by the Fermi 
velocity $v_F$.   

In the leading approximation, quasiparticle  excitations in the vicinity of the 
nodal points (hereafter referred to as valleys) can be described by the Dirac Hamiltonian \cite{semenoff} 
\be
H=v_F[{\hat 1}\otimes{\hat \sigma}_xk_x+{\hat \tau}_z\otimes{\hat \sigma}_yk_y] 
\ee
where ${\hat \sigma}_i$ is the triplet of the Pauli matrices acting in  the 
space of spinors $\Psi=(u_A,v_B)$ composed of the values of the electron wave 
function on the $A$ and $B$ sublattices of the hexagonal lattice of graphene, 
whereas the triplet ${\hat \tau}_i$ acts in the valley subspace. In the absence 
of magnetic field, the Hamiltonian (1) remains a unity matrix in the physical spin subspace.  

In the presence of disorder, the quasiparticles experience both intra- and 
inter-valley scattering. Upon averaging over disorder, the most general form 
of the elastic four-fermion vertex induced by disorder can be described by the expression
\bea 
{\hat W}=g_4[{1+{\tau}_z\over 2}\otimes{1+{\tau}_z\over 2}+ {1-{\tau}_z\over 2}
\otimes{1-{\tau}_z\over 2}]\nonumber\\ +g_2[{1+{\tau}_z\over 2}\otimes{1-
{\tau}_z\over 2}+ {1-{\tau}_z\over 2}\otimes{1+{\tau}_z\over 2}]
\nonumber\\
+g_1[{{\tau}_+}\otimes{{\tau}_-}+{{\tau}_-}\otimes{{\tau}_+}] +g_3[{{\tau}_+}
\otimes{{\tau}_+}+{{\tau}_-}\otimes{{\tau}_-}] 
\eea 
where we used the customary "g-oloqical" notations to denote the processes of 
forward scattering between fermions from the same ($g_4$)  and different ($g_2$) 
valleys, as well as those of "backward" scattering ($g_1$)  with the momentum 
transfer close to ${\vec K}_1-{\vec K}_2$. Besides, we have included the 
possibility of "umklapp" scattering ($g_3$), in which case the total momentum 
of two fermions changes by $2({\vec K}_{1}-{\vec K}_2)={\vec K}_{2}-{\vec K}_1+{\vec Q}$. 
A justification for this
(not immediately obvious, considering that the total momentum changes by only a 
fraction of the reciprocal latice vector) extension of 
the disorder model will be discussed below.

In the framework of the self-consistent Born approximation (SCBA), the effect 
of disorder on the quasiparticle spectrum is described by a self-energy which obeys the equation 
\be 
{\hat \Sigma}^R(\omega,{\vec k})=\int {d{\vec q}\over (2\pi)^2}Tr{\hat W}
{\hat G}^R(\omega,{\vec k}+{\vec q})
\ee 
where the retarded Green function is given by the expression 
\be 
{\hat G}^R(\omega,{\vec k})={(\omega-\mu+\Sigma^R){\hat 1}\otimes{\hat 
1}+v_F({\hat 1}
\otimes{\hat \sigma}_xk_x+ {\hat \tau}_z\otimes{\hat \sigma}_yk_y)
\over {(\omega-\mu+\Sigma^R)^2-v^2_Fk^2}} 
\ee 
that includes a chemical potential $\mu$ which allows one to account for a variable electron  
density. 

Notably, the solution to Eq.(3) turns out to be diagonal in the valley subspace,
${\hat \Sigma}^R(0,0)=i\gamma{\hat 1}\otimes{\hat 1}$, and proportional to
the inverse elastic lifetime 
\be
\gamma={\pi\nu_F\over 2}(g_4+g_1) 
\ee 
where $\nu_F$ is the density of states at the Fermi energy. In the low-doping 
limit ($\mu<\gamma$), a finite value of $\nu_F=-\int d{\vec k}/(4\pi^3)Im G(\mu,{\vec k})$ 
is dominated by disorder. A similar phenomenon has been extensively studied in the context of normal 
quasiparticle transport in dirty $d$-wave superconductors 
where $\mu=0$ and the spectrum possesses an exact particle-hole symmetry \cite{lee}. 
In both cases one obtains the self-consistent DOS as $\nu_F=(\gamma/2\pi)\ln(v_F/a\gamma)$.  

However, most of the data of Refs.\cite{geim,kim} pertain to the regime of
relatively high dopings ($\mu>>\gamma$), as indicated by, e.g., the measured mean free path, which was found 
to be of order $\sim 1\mu m$ at electron densities $n_e\sim 10^{13}cm^{-2}$. 
In this regime, the value of $\nu_F=k_F/(2\pi v_F)$ is controlled by the finite radius of the 
Fermi surface $k_F=\mu/v_F=(4\pi n_e)^{1/2}$ and is only weakly affected by disorder. 

Therefore, the double-pole Green function (4) given by a four-by-four matrix 
can be well approximated by a pair of two-by-two matrices containing single-poles 
\be 
{\hat G}_{1,2}(\omega,{\vec k})\approx{1+{\hat \sigma}_x\cos\phi_k\pm
{\hat \sigma}_y\sin\phi_k\over {2(\omega-\xi_k+i\gamma)}} 
\ee 
where $\xi_k=v_Fk-\mu$ and $\phi_k=\tan^{-1}k_y/k_x$.  

Even more importantly, the presence of a large parameter $\mu/\gamma$ facilitates a 
systematic account of quantum interference corrections to the SCBA results. 
We mention, in passing, that in the low-doping ($\mu<\gamma$) limit the only 
parameter that can (at least, in principle) be used for the analysis of such 
corrections is the (inverse)  number of valleys, whose actual value is, of 
course, $N_v=2$. The same caveat plagues the analysis of dirty $d$-wave 
superconductors where the strength of the conductivity corrections is governed by the 
number of pairs of opposite nodes of the order parameter, whose physical value is
again equal two.   

Luckily, in the high-density regime that, so far, has been probed in graphene, the 
leading quantum correction to the zeroth-order (Drude) conductivity stem
from the standard single-Cooperon (fan-shaped) diagram. From the technical standpoint, however, the calculation 
of the corresponding correction appears to be somewhat more involved due to the 
matrix structure of the  Green function (6) and the vertex (2).  

The Drude conductivity itself is given by the standard expression $\sigma_0=(e^2/h)
\Gamma\mu/\gamma$ where the renormalization factor $\Gamma$ accounts for a
 ladder series of vertex corrections associated with one of the two current 
 operators ${\vec J}=v_F(\cos\phi_k,\sin\phi_k)$ inserted into the fermion 
 loop in the diagrammatic representation of the Kubo formula.  
Being given by a non-singular (angular momentum $m=1$) diffusion mode in the 
expansion over the angular harmonics $e^{im\phi_k}$, $\Gamma$ is a function of the parameters $g_i$.  

Expanding the expression for the Cooperon in the same basis as that used in 
Eq.(2), we obtain equations for the corresponding amplitudes $C_i$, each of 
which is a matrix in the direct product of two ${\hat \sigma}_i$ subspaces   
\bea 
C_1=g_1+g_2H_{12}C_1+g_1H_{21}C_2\nonumber\\ 
C_2=g_2+g_2H_{12}C_2+g_1H_{21}C_1\nonumber\\ 
C_3=g_3+g_4H_{11}C_3+g_3H_{22}C_4\nonumber\\ 
C_4=g_4+g_4H_{11}C_4+g_3H_{22}C_3 
\eea 
where $H_{ij}(\omega,{\vec q})=\int {d{\vec k}\over (2\pi)^2}G^R_i
(\epsilon+\omega/2,k+q/2) G^A_j(\omega/2-\epsilon,q/2-k)$ denotes a convolution of a pair of Green functions. 
Computing the latter at $\omega,v_Fq<<\gamma$, we obtain 
\bea 
{\hat H}_{11,12}(\omega,q)={\hat H}_{22,21}(\omega,q)={\pi\nu_F\over 4\gamma}
[(1+{i\omega\over 2\gamma}-{v_F^2q^2\over 16\gamma^2}){\hat 1}\otimes{\hat 1}\nonumber\\
 -{1\over 2}(1+{i\omega\over 2\gamma}-{v_F^2q^2\over 16\gamma^2}{3\cos^2\phi_k+\sin^2
 \phi_k\over 4}){\hat \sigma_x}\otimes{\hat \sigma_x}\nonumber\\ 
 \mp {1\over 2}(1+{i\omega\over 2\gamma}
 -{v_F^2q^2\over 16\gamma^2}{\cos^2\phi_k+3\sin^2\phi_k\over 4}){\hat \sigma_y}\otimes{\hat \sigma_y}]
\eea 
It can be readily seen that Eqs.(7) split onto two pairs which only couple $C_{1,2}$ and 
$C_{3,4}$, respectively. Their solutions read 
\bea {\hat C}_{1,2}(\omega,q)={2\gamma c_{1,2}\over \pi\nu_F}{{\hat 1}\otimes{\hat 1}- 
{\hat \sigma_x}\otimes{\hat \sigma_x}+{\hat \sigma_y}\otimes{\hat \sigma_y}+ 
{\hat \sigma_z}\otimes{\hat \sigma_z}\over  {(g_4-g_2)/(g_2+g_1)+v_F^2q^2/16\gamma^2-i\omega/2\gamma}}\nonumber\\ 
{\hat C}_{3,4}(\omega,q)={2\gamma c_{3,4}\over \pi\nu_F}{{\hat 1}\otimes{\hat 1}-
{\hat \sigma_x}\otimes{\hat \sigma_x}-{\hat \sigma_y}\otimes{\hat \sigma_y}- 
{\hat \sigma_z}\otimes{\hat \sigma_z}\over  {(g_1-g_3)/(g_3+g_4)+v_F^2q^2/16\gamma^2-i\omega/2\gamma}}\ 
\eea 
where $c_1=(g^2_1+g_1g_4)/(g_4-g_2+2g_1)(g_1+g_2)$, $c_2=(g_2g_4+g_1g_2-g^2_2+g^2_1)/(g_4-g_2+2g_1)(g_1+g_2)$,  
$c_3=(g_3g_1+g_3g_4)/(g_1+g_3)(g_3+g_4)$, and $c_4=(g_1g_4+g^2_3)/(g_1+g_3)(g_3+g_4)$.  

The quantum conductivity correction (including spin) 
\be 
\delta\sigma_{xx}=-{e^2\over h}{\pi\nu_Fv_F^2\over 16\gamma^3}
\int {d{\vec q}\over (2\pi)^2}Tr[{\hat C}_1(0,q)+{\hat C}_4(0,q)] 
\ee 
involves the $C_1$ and $C_4$ components of the Cooperon, whose contributions 
turn out to be negative and positive, respectively.  The logarithmic temperature-dependent 
part of Eq.(10) can be cast in the form 
\be 
\delta\sigma_{xx}={2e^2\over {\pi h}}
\ln{max[\Gamma_\phi/\gamma,|g_4-g_2|/(g_1+g_2)]^{c_1}\over 
{max[\Gamma_\phi/\gamma,|g_1-g_3|/(g_4+g_3)]^{c_4}}} 
\ee 
where we introduced an inelastic phase relaxation rate 
$\Gamma_\phi(T)$ which provides a cutoff in the momentum 
integration and diverges at $T\to 0$.  

The analysis of Eq.(11) reveals that, in the absence of a fine tuning between 
the amplitudes of backward scattering and umklapp processes ($g_1\neq g_3$) 
the $C_4$ Cooperon always acquires a gap $\sim \gamma|g_1-g_3|/(g_4+g_3)$. On the other hand, the $C_1$ mode
remains gapless, provided that all the forward scattering processes are controlled 
by the same amplitude (i.e., $g_2=g_4$).

Conversely, making the $C_1$ mode gapful and inverting the sign of the 
conductivity correction would only be possible under the condition 
$|g_1-g_3|<<|g_2-g_4|$ which is unlikely to be satisfied for any realistic 
impurity potential that yields equal amplitudes of the processes of intra- and 
inter-valley forward scattering.  

The antilocalizing behavior predicted in Ref.\cite{ando} for $g_{1}<<g_2=g_4$
and $g_3=0$ ($c_1=1/2,~~ c_4=1$) can only occur at intermediate temperatures
(namely, at $\gamma g_1/g_2<\Gamma_\phi(T)<\gamma$), whereas
at still lower temperatures the overall sign of (11) reverts to negative,
thereby suggesting the onset of rather conventional weak localization.  

The above conclusions apply to the general disorder model (2). 
However, in the situation where disorder is realized as a random 
distribution of impurities with a concentration $n_i$ and a (short-range) 
potential $u(q)$, it suffices to introduce only two independent parameters
\bea
g_2=g_4={n_iu^2(0)\over {1+(\pi\nu_Fu(0))^2}}\nonumber\\
g_1=g_3={n_iu^2({\vec K}_1-{\vec K}_2)\over {1+(\pi\nu_Fu
({\vec K}_1-{\vec K}_2))^2}}
\eea
The above expressions represent a $\hat T$-matrix computed for 
an arbitrary strength of disorder, the customary Born and unitarity 
(where the scattering phase approaches $\pi/2$) limits corresponding to $u\to 0$ and $\infty$, respectively.
[In the case of a genuine long-range (unscreened) impurity potential, 
the dependence $u(q)$ on the transferred momentum makes the explicit formulas
for $g_i$ more involved, though.] 

Provided that the relations (12) between the 
parameters $g_i$ hold, one obtains $c_1=c_4=1/2$, and the logarithmic 
term in (11) vanishes as a result of the exact cancellation between the 
contributions of the localizing ($C_1$) and antilocalizing ($C_4$) 
Cooperon modes. It has to be stressed, however,
that such a strong suppression of (anti)localization 
would only be possible due to an opening of the umklapp channel. In turn, 
the latter requires an emergence of a crystal superstructure with the 
wave vector $2({\vec K}_1-{\vec K}_2)=(2/3)({\vec Q}_1+{\vec Q}_2)$ (or equivalent). 

While an isolated sheet of weakly-interacting graphene would apparently lack 
such a superstructure, it is conceivable that the latter might emerge if a 
commensurate substrate were used during the process of microfabfication. 
Besides, a commensurate corrugation could occur \cite{shikin}
due to the Coulomb correlations that can induce spatially periodic patterns of the electron density itself.
The possibility of a spontaneous formation of such charge density 
wave states has long been discussed in the general context of degenerate 
semimetals and, specifically, in graphene \cite{cdw}. 

Next, we comment on the technical details presented in Ref.\cite{ando} where the first 
prediction of the antilocalization behavior for $g_2=g_4$ and 
$g_1=g_3=0$ was made. In essence, instead of working directly with 
the matrices in the sublattice subspace, the authors of Ref.\cite{ando} 
used single-component (projected) Green functions $G(\omega,{\vec k})=1/(\omega-\xi_k+i\gamma)$ 
complemented by matrix elements of a (purely forward-scattering) impurity potential 
between different Bloch wave functions. 

To that end, in Ref.\cite{ando} the latter were chosen 
in a particular (asymmetrical) gauge 
\bea
\psi_{1k}={1\over {\sqrt 2}}\pmatrix{\pm 1, & e^{i\phi_k}, & 0, & 0},\nonumber\\
\psi_{2k}={1\over {\sqrt 2}}\pmatrix{0, & 0, & e^{i\phi_k}, & \pm 1}
\eea 
The aforementioned matrix element of the impurity potential
is then given by a complex-valued 
expression $<{\vec p}|u|{\vec k}>\propto(1+e^{i(\phi_k-\phi_p)})$.

Next, the authors of Ref.\cite{ando} asserted that the equations for the 
Cooperon mode involve the disorder-induced vertex  
\bea
{W}_{k,-k,p,-p}=n_i<{\vec p}|u|{\vec k}><{-\vec p}|u|{-\vec k}>\nonumber\\
\propto e^{i(\phi_k-\phi_p)}[1+\cos({\phi_k-\phi_p})]
\eea
which describes scattering from the Bloch states with momenta ${\vec k}+{\vec q/2}$ and $-{\vec k}+{\vec q/2}$
into ${\vec p}+{\vec q/2}$ and $-{\vec p}+{\vec q/2}$ in the limit $q\to 0$. 

It was then argued in Ref.\cite{ando} that the full Cooperon amplitude 
inherits the phase factor from Eq.(14) and, therefore, becomes negative for 
$\vec k\approx -{\vec p}$ where $e^{i(\phi_k-\phi_p)}\approx -1$, which 
configuration of the momenta provides a dominant contribution to the weak
 localization correction. Thus obtained, the negative sign of the Cooperon 
 amplitude was claimed to be instrumental for the onset of antilocalizing behavior. 
 
The validity of the above argument can be disputed by observing that 
the choice of a different (symmetrical) gauge for the Bloch wave functions
\bea
\psi^\prime_{1k}={1\over {\sqrt 2}}\pmatrix{\pm e^{-i\phi_k/2}, & e^{i\phi_k/2}, & 0, & 0},\nonumber\\
\psi^\prime_{2k}={1\over {\sqrt 2}}\pmatrix{0, & 0, & e^{i\phi_k/2}, & \pm e^{-i\phi_k/2}}
\eea
would result in a purely real matrix element $<p|u|k>\propto 2\cos{\phi_k-\phi_p\over 2}$, the use of which yields
Eq.(14) without the said phase factor, hence the (deemed to be crucially important) 
sign change of the Cooperon amplitude does not seem to occur. 

Although the observed gauge dependence of the Cooperon is perfectly consistent 
with its being a gauge non-invariant two-particle amplitude, the conductivity 
correction is, of course, supposed to be gauge invariant. A resolution of such 
apparent contradiction goes as follows. 

In fact, the vertex that ought to be used in the construction of the Cooperon is 
the particle-hole (exchange) amplitude ${W}_{k,p,p,k}$ which is manifestly 
gauge invariant and given by Eq.(14) without the phase factor in question. In this 
regard, it is important to realize that the amplitudes ${W}_{k,-k,p,-p}={W}_{q,q-k,p,q-p}|_{q\to 0}$ 
and ${W}_{k,p,p,k}|_{k\to -p}$ represent two non-commuting (and, in this particular case, unequal) 
limits of the general vertex (2).

On a side note, it might also be tempting to try to explain the experimentally found 
suppression of localization by the presence of the 
factor $[1+\cos({\phi_k-\phi_p})]$ vanishing for ${\vec k}=-{\vec p}$ in Eq.(14). 
Notice, however, that the Cooperon built out of such a vertex lacks this factor, 
since the gapless pole develops for only one (namely, $m=0$) angular harmonic, whereas the other two harmonics
that constitute the above angular-dependent factor ($m=\pm 1$)
receive only a finite enhancement by a factor of two. 

Thus, an obvious suppression of the backward scattering off
of an individual impurity (which fact has often been mentioned in the context of electron 
transport in 1D carbon nanotubes) does not necessarily imply the absence of (anti)localization 
in the deep diffusive regime in 2D graphene. 

The true origin of the antilocalizing behavior exhibited by Eq.(11)
for $|g_1-g_3|<<|g_2-g_4|$, can be traced back
to the negative signature of the expansion of the Cooperon mode $C_4$ over the product 
basis $\sigma_a\otimes\sigma_b$, as opposed to that of $C_1$ (see Eqs.(9)). Interpreting 
the sublattice index as a fictitious spin one-half (${\vec S}_{1,2}={\vec \sigma}_{1,2}/2$), 
one can associate the $C_4$ component with the singlet mode
${\hat C}_4\propto ({\hat 1}\otimes{\hat 1}-{1\over 2}({\vec S}_1+{\vec S}_2)^2$)
which, in the spin-orbit context, is known to be a common source of antilocalization.

In this regard, the present anilocalizing behavior is similar to that predicted in the case 
of 2DEG with a strong spin-orbit coupling of either Rashba or Dresselhaus kind \cite{skvortsov}
\be
H_{so}= \alpha({\hat \sigma}_xk_y-{\hat \sigma}_yk_x)+\beta({\hat \sigma}_xk_x-{\hat \sigma}_yk_y)
\ee
The details of the calculations presented above differ from those pertaining to the case of 
the spin-orbit coupling (for one thing, except for a splitting, the 
linear in momentum SO-coupling does not alter the fermion dispersion,
whereas in the case of graphene it is the linear term which is solely 
responsible for the dispersion). 

Yet another argument invoked by the authors 
of Ref.\cite{ando} to support their prediction of 
the antilocalizing behavior intermediate temperatures  
exploits the notion of the Berry 
phase associated with the electron wavefunctions (13). 

Indeed, the first Chern number ("vorticity")
\be
\Phi=i\oint <\psi_k|{\vec \nabla}_k|\psi_k>,
\ee
where the momentum space integral is taken along any contour enclosing 
one of the Dirac points ${\vec K}_{1,2}$, yields $\Phi_{1,2}=-\pi$ for both wavefunctions (13). 
In Ref.\cite{ando} (see also \cite{blioch}) this observation was used to assert that the 
interference between any pair of counter-propagating electron waves becomes destructive, 
thereby giving way to the antilocalizing behavior.

However, the Berry phase (17) would be identically  
zero, if the wave functions (14) were used instead of (13), which 
suggests that its relationship with the Cooperon
is somewhat more subtle.
To that end, it is worth mentioning that the two Dirac species emerging 
in the continuous (low-energy) description of the original lattice system
must have opposite chiralities \cite{kane}, as required for compliance with the ubiquitous Nielsen-Ninomiya 
("fermion doubling")
theorem \cite{doubling}. In turn, this observation seems to suggest that, even if the Berry phases of 
the individual species happen to be finite ($\Phi_{1,2}=\pm\pi$), they should still add up to zero.

A cancellation of the overall Berry phase would also be consistent with the vanishing 
of the quantum conductivity correction in the presence of equally strong intervalley backward 
and umklapp scattering, which, in effect, make the two Dirac points indistinguishable (the 
shortest reciprocal lattice vector becomes equivalent to ${\vec K}_1-{\vec K}_2$). 

Recently, the role of the Berry phase has also been brought up in the context of the spin Hall effect 
generated by a spin-orbit coupling. 
Interestingly enough, the dependence of the spin Hall conductivity upon 
the coefficients $\alpha$ and $\beta$ from Eq.(16) (and, especially,
its exact vanishing at $\alpha=\pm\beta$) can be established 
with the use of the Bloch wave functions in either asymmetrical (similar to Eq.(13)) or symmetrical (similar to Eq.(14)) gauge,
where the corresponding Berry phase is non-zero and zero, respectively
(cf., e.g., the two references in \cite{shen}).
This observation, too, indicates that the relationship between
the Berry phase and transport properties might be rather intricate.

Before concluding, a few more comments are in order. 

In reality, the weak (anti)localization corrections are accompanied by the Aronov-Altshuler ones, 
which stem from the interference between disorder and Coulomb interactions. Despite a similar 
temperature dependence ($\delta_{AA}\sigma_{xx}=-(e^2/\pi h)\ln(\gamma/T)$), the former, 
but not the latter, gets quenched by an applied magnetic field, thus allowing one to discriminate between the two.

Also, considering the potential attainability of the very low-density regime, it would be of 
interest to extend the analysis of the localization effects to the case $\mu<\gamma$. However, 
as we have already pointed out, in this regime a systematic expansion of the conductivity 
corrections would only be possible for an (unphysically) large number of valleys. 

Moreover, even if this number were indeed large, the calculation itself would still pose a significant challenge. 
The main technical difficulty here
is due to the double-pole structure of the 
Green functions (4), which results in the emergence of additional gapless Cooperon and diffusion modes
in the RR (AA) channel, alongside the conventional (RA) one. 
A similar situation has been encountered in the context of 
dirty $d$-wave superconductors \cite{altland,ayash1} 
where it has been argued that 
the quasiparticle conductivity receives aditional logarithmic contributions from the processes 
involving RR (AA) Cooperons \cite{china}. 

Furthermore, the authors of Ref.\cite{china} found the overall logarithmic correction to 
the spin (thermal) conductivity
of normal $d$-wave quasiparticles to be negative, contrary to 
(in this case, ill-defined due to a non-conservation of charge) 
electrical one.
In view of the formal differences between the Dirac-like descriptions of planar $d$-wave 
superconductors and graphene, it is, at least, conceivable that the correction to the electrical 
conductivity of graphene could change sign in the low-density regime, too.

Lastly, we comment on the possiblity (as well as the need) 
of including other types of randomness. While the vertex (2) is devised as a general model of 
short-range (screened) impurities generating a random scalar potential, it obviously 
misses out on those types of disorder that can be best represented by either a random 
vector potential or a random mass of the Dirac fermions. 

The latter might be relevant in the situation where a spatially inhomogeneous charge or spin 
density wave ordering sets in. 
In turn, the former can be utilized to describe extended lattice defects (dislocations, disclinations, 
and cracks) in terms of a random magnetic flux \cite{guinea}. In Ref.\cite{guinea}, it was assumed that 
the corresponding random vector potential (rather then the 
flux) is $\delta$-correlated in space
($<{\vec A}_q{\vec A}_{-q}>=const$). Although this model possesses a conformal invariance 
and, therefore, can be analyzed in quite some detail \cite{altland,ludwig}, a more adequate to the task would obviously 
be the model
of a long-range correlated random vector potential
($<{\vec A}_q{\vec A}_{-q}>\propto 1/q^2$).
The latter is known to show a rather different behavior
even in the high-energy (ballistic) regime, as demonstrated in Ref.\cite{ayash2}.

A further investigation into these and related issues will be presented elsewhere.

In summary, we carried out a comprehensive analysis of the first-order quantum correction to the 
conductivity of doped graphene. 
We identified the conditions under which the conductivity correction 
becomes positive, negative, or zero.
The earlier analysis of Ref.\cite{ando}
has been critically assessed, and the open problems have been outlined.

This research was supported by NSF under Grant DMR-0349881.  The author acknowledges 
valuable communications with E. Abrahams, A. Geim, V. Gusynin,
and H. Suzuura.


\begin{thebibliography}{99}
 
\bibitem{geim} K. S. Noveselov et al, Science {\bf 306}, 666 (2004); Nature {\bf 438}, 197 (2005).

\bibitem{kim} Y. Zhang et al, Nature {bf 438}, 201 (2005);
Appl. Phys. Lett.{\bf 86}, 073104 (2005);
Phys. Rev. Lett. {\bf 94}, 176803 (2005).

\bibitem{semenoff} G. Semenoff, Phys. Rev. Lett.{\bf 53}, 2449 (1984); 
F. D. M. Haldane, ibid {\bf 61}, 2015 (1988); E. Fradkin, Phys. Rev. {\bf B33}, 3257 (1986);
J.Gonzalez, F.Guinea, and M.A.H.Vozmediano,
Nucl.Phys.{\bf 406}, 771 (1993); ibid {\bf B424}, 595 (1994).

\bibitem{gusynin} V. P. Gusynin and S. G. Sharapov, 
Phys. Rev. Lett.{\bf 95}, 146801 (2005); 
cond-mat/0512157; N.M.R. Peres, F. Guinea, and A. H. Castro-Neto,
cond-mat/0506709,0512091; E. McCann and V.I. Falko,
cond-mat/0510237.  

\bibitem{kopelevich}  G.P. Mikitik and Y.V. Sharlai,
Phys. Rev. Lett. {\bf 82}, 2147 (1999);
I.A. Lukyanchuk and Y. Kopelevich, 
ibid {\bf 93}, 166402 (2004); Gusynin and S. G. Sharapov, Phys. Rev. {\bf
B71}, 125124 (2005).

\bibitem{guinea} J. Gonzalez, F. Guinea, and M. A. H. Vozmediano,
Phys. Rev. Lett. {\bf 77}, 3589 (1996); 
Phys. Rev. {\bf B59}, 2474 (1999); ibid {\bf B63}, 134421 (2001);
T. Stauber, F. Guinea, and M. A. H. Vozmediano, cond-mat/0311016;
M.A.H.Vozmediano et al, cond-mat/0505557;
N.M.R. Peres, F. Guinea, and A. H. Castro-Neto,
cond-mat/0507061;0512091; J. Nilsson et al,
cond-mat/0512360.  

\bibitem{dvk} D. V. Khveshchenko, Phys. Rev. Lett. {\bf 87}, 206401 (2001); 
ibid {\bf 87}, 246802 (2001); E. V. Gorbar, V. P. Gusynin, V. A. Miransky, 
and I. A. Shovkovy, Phys. Rev. {\bf B66}, 045108 (2002); 
D. V. Khveshchenko and H. Leal, Nucl. Phys. {\bf B687}, 323 (2004);
D. V. Khveshchenko and W. F. Shively, cond-mat/0510519.

\bibitem{ando} H. Suzuura and T. Ando,
Phys. Rev. Lett. {\bf 89}, 266603 (2002);
J. Phys. Soc. Jpn. {\bf 72}, 69 (2003).

\bibitem{lee} P.A. Lee, Phys. Rev. Lett.{\bf 71}, 1887 (1993);
A.C. Durst and P.A. Lee, Phys. Rev. {\bf B62}, 1270 (2000).

\bibitem{shikin} A. M. Shikin et al, Phys. Rev. Lett.{\bf 90}, 256803 (2003).

\bibitem{cdw} 
A.L.Tchougreeff and R.Hoffmann, J. Phys. Chem.{\bf 96}, 8993 (1992);
F.R.Wagner and M.B.Lepetit, ibid {\bf 100}, 11050 (1996).

\bibitem{skvortsov} M.A. Skvortsov, JETP Lett.{\bf 67}, 133 (1998);
I.V. Gornyi, A.P.Dmitriev, and V.Y. Kachorovskii,
ibid {\bf 68}, 338 (1998); L.E. Golub, cond-mat/0412047.

\bibitem{blioch} K.Y. Blioch, Phys. Lett. {\bf A344}, 127 (2005).

\bibitem{kane} C.L.Kane and E.J. Mele, Phys Rev. Lett.
{\bf 95}, 146802 (2005).

\bibitem{doubling} H.B. Nielsen and M. Ninomiya, Nucl. Phys.{\bf B185}, 20 (1981); ibid {\bf B193}, 173 (1981).

\bibitem{shen} S.-Q. Shen, Phys. Rev. {\bf B70}, 081311 (2004);
N.A. Sinitsyn et al, ibid {\bf 70}, 081312 (2004).

\bibitem{altland} A.Altland and M.R.Zirnbauer, Phys.Rev.{\bf B55}, 1142 (1997); 
A. Altland, B. D. Simons, and M. R. Zirnbauer, Phys. Rep. {\bf 359}, 283 (2002).

\bibitem{ayash1} D.V. Khveshchenko, A.G. Yashenkin, and I.V. Gornyi,
Phys. Rev. Lett. {\bf 86}, 4668 (2001); A.G. Yashenkin et al,
ibid {\bf 86}, 5982 (2001).  

\bibitem{china} Y.A.Yang et al,  
Euro. Phys. Lett. {\bf 63}, 111 (2003);
Comm. Theor. Phys. {\bf 42}, 309 (2004).

\bibitem{ludwig} A. W. W. Ludwig, M. P. A. Fisher, R. Shankar, 
and G. Grinstein, Phys. Rev.{\bf 50}, 7526 (1994).

\bibitem{ayash2} D. V. Khveshchenko and A. G. Yashenkin, Phys. Lett. {\bf A309}, p.363 
(2003); Phys. Rev.{\bf B67}, 052502 (2003). 
 
\end{thebibliography}
\end{document}